\documentclass[a4paper,12pt]{article}
\usepackage{amsmath,amsfonts,amssymb}

\newtheorem{twr}{Theorem}
\newtheorem{defi}{Definition}

\newcommand{\ket}[1]{\mbox{$| #1 \rangle$}}

\newcommand{\kb}[2]{\ensuremath{| #1 \rangle\!\langle #2 |}}
\newcommand{\tr}{\mbox{$\mathrm{tr}$}}

\begin{document}

\title{Classical nonintegrability of a quantum chaotic $SU(3)$ Hamiltonian system}

\author{Adam Sawicki$^{1,2}$\footnote{email: assawi@interia.eu} and
Marek Ku\'s$^2$\footnote{email: marek.kus@cft.edu.pl}
\\
\\
$^1$Department of Physics, University of Warsaw, \\
ul.\ Ho\.za 69, 00-667 Warszawa, Poland,
\\
\\
$^2$Center for Theoretical Physics, Polish Academy of Sciences, \\
Al.\
Lotnik\'ow 32/46, 02-668 Warszawa, Poland}

\maketitle

\abstract We prove nonintegrability of a model Hamiltonian system defined on
the Lie algebra $\mathfrak{su}_3$ suitable for investigation of connections
between classical and quantum characteristics of chaos.

\medskip



\noindent MSC 53D17, 70H07, 12H05

\section{Introduction}

The problem of distinguishing on the quantum level between integrability and
chaos of classical systems is a recurrent, if not the principal, topic of the
quantum chaos theory. The usual definitions or signatures of the chaotic
motion based on such phase-space notions like Lyapunov exponents or various
degrees of ergodicity lack sense on the quantum level due to the absence of
the proper notion of the phase space in quantum mechanics. Instead, in search
of criteria of quantum chaos one should resort to purely quantum
characteristics of a system, like e.g.\ its spectral properties.

A quarter of a century ago, Bohigas and coworkers \cite{bohigas84} proposed a
characterization of quantum chaos based on the statistical theory of spectra.
According to their hypothesis most quantum systems whose classical limit is
chaotic display universal spectral fluctuations determined by the Random
Matrix Theory (RMT) \cite{mehta91}. Generic classically nonintegrable systems
exhibit thus level repulsion i.e.\ the probability of finding two adjacent
energy levels tends to zero with the difference between their energies.

A vast numerical and experimental evidence [3, 4] in favor of this hypothesis
was collected during last twenty years. At least two general strategies
providing theoretical arguments supporting the Bohigas-Giannoni-Schmidt
conjecture were developed. The first one employs statistical mechanics of a
fictitious gas of eigenvalues undergoing parametric dynamics, where the role
of time is played by a control parameter controlling the transition from an
integrable to a chaotic system \cite{pechukas83,yukawa85}. The other approach
takes its roots in the semiclassical quantization \textit{via} classical
periodic orbits pioneered by Gutzwiller \cite{gutzwiller90} and extends ideas
from the theory of disordered systems to dynamical systems \cite{haake00}. In
both approach, at some stage one invokes some statistical hypothesis which
justifies ascribing statistically inferred properties to an individual system
which is the object of actual numerical or experimental examinations. On the
other hand the  proofs of nonintegrability on the classical level are usually
based solely on numerical investigations. We do not know of any example of a
model system for which there exist analytical proofs of the classical
nonintegrability on one side and of the repulsion between quantum levels on
the other. The aim of the presented investigations is to provide such a
model. In the present paper we will concentrate on the classical side of the
problem and show its classical nonintegrability.

In a series of papers \cite{gk98,ghk00,sk06} one of us proposed a class of
models taking their origins in atomic physics and quantum optics in which
dynamical variables were elements of a compact semisimple Lie algebra in some
particular irreducible representation. The classical limit was attained by
going with the dimension of the representation to infinity. For Lie algebras
with the rank larger that one (e.g.\ for $\mathfrak{su}_3$ algebra) there are
more than one, `natural' ways of performing this limiting procedure. In the
effect there are several inequivalent classical corresponding classical
systems, differing e.g.\ by the dimensionality of the classical phase space.
The interplay between the number of degrees of freedom and the dimensionality
of the space is crucial for the (non-)integrability of a classical
Hamiltonian system. A particular quantum system can be classically integrable
or not, depending on the way the classical limit is approached. The story
becomes interesting if, basing on the above observation, one can say
something about purely quantum features (e.g.\ spectral properties) of the
quantum system in question. The affirmative reply based on numerical
investigation of spectra and classical characteristics of chaos, was given in
\cite{ghk00}.

In the present paper we examine again the above models in order to prove in
an analytical, rather than numerical manner the nonintegrability. Admittedly
it is a rather minimalistic goal. Nobody claims that mere nonintegrability of
the classical system is sufficient for the repulsion of the quantum energy
levels, we believe that the Bohigas-Giannoni-Schmidt conjecture gives correct
predictions only for `sufficiently chaotic' systems. Nevertheless the
nonintegrability is a necessary prerequisite and we decide to pursue such a
modest goal of showing it for the considered system.

\section {Nonintegrability of Hamiltonian systems}\label{sec:HS}
In this section we present a brief introduction to the Morales-Ramis theory
of non-integrability of Hamiltonian systems \cite{morales99,audin08}. The aim
is to give some basic definitions and theorems (without proofs) and to
present general scheme which should be followed when one wants to prove
non-integrability. The whole concept is based on the differential Galois
theory (here good references are \cite{kaplansky57,kolchin73,pommaret83}), so
we also give some key ideas of it.

\subsection {Differential Galois theory}
\label{subsec:DG} Differential Galois theory (a.k.a.\ Picard-Vessiot theory)
is an analogue of the classical Galois theory (which deals with algebraic
equations), for linear differential equations. The main object of the theory
is here a differential field $K$ i.e.\ an algebraic field equipped with a
differentiation, i.e.\ a linear mapping $^\prime :K\rightarrow K$ satisfying
the Leibniz rule, $(fg)^\prime=f^\prime g+fg^\prime$.

The object of our interest is a linear differential equation:
\begin{equation}\label{DG1}
y^{(n)}+a_{n-1}y^{(n-1)}+\ldots+a_1y^{\prime}+a_0y=0,
\end{equation}
where the coefficients $a_i$ are elements of $K$ (it is good to think
$K=C(x)$, the field of rational functions). The natural question which arises
in the differential Galois theory is: can one solve the equation (\ref{DG1})
using some ``elementary'' operations? Of course usually the solution of
(\ref{DG1}) is not contained in the field $K$ and we have to extend $K$ to a
larger one. The smallest extension of $K$ which contains all solutions of
(\ref{DG1}) is called the Picard-Vessiot extension. It exists always when the
field of constants of $K$ is algebraically closed \cite{kaplansky57}.

To be more precise we have to define what we mean by "elementary" operations.
\begin{defi}
Let $K$ be a differential field. We say that the equation (\ref{DG1}) is
solvable in the Liouville functions category when the Picard-Vessiot
extension of $K$ can be obtained from $K$ in a finite number of steps each
one being an extension of $K$ by adding:
\begin{itemize}
\item[1.] $\alpha _i$ which is algebraic over $K$,
\item[2.] $\alpha _i$ which is such that $\alpha _i^\prime \in K$,
\item[3.] $\alpha _i$ which is such that $\frac{\alpha _i^\prime}{\alpha
    _i}\in K$.
\end{itemize}
\end{defi}
The steps 1., 2., and 3. correspond to the operations of taking roots of
polynomial equations, integration, and taking the exponent of an integral ---
the natural operations when one solves a differential equation. The last
important definition is that of the differential Galois group:
\begin{defi}
Let $L\supset K$ be a Picard-Vessiot extension of $K$ for the equation
(\ref{DG1}). The differential Galois group $Gal(L\supset K)$ of the extension
$L\supset K$ is the group of all differential automorphisms (algebraic
automorphisms that commute with the differentiation) of $L$ which are
identity on $K$.
\end{defi}
Let $y_1$ be a solution of (\ref{DG1}) and $\sigma\in Gal(L\supset K)$. Then:
\begin{eqnarray}
\sigma(y_1^{(n)}+a_{n-1}y_1^{(n-1)}+\ldots+a_1y_1^{\prime}+a_0y_1)=\sigma(0)=0
\end{eqnarray}
Using properties of $\sigma$, namely the fact that it is a differential
automorphism, we obtain:
\begin{eqnarray}
\sigma(y_1)^{(n)}+a_{n-1}\sigma(y_1)^{(n-1)}+\ldots+a_1\sigma(y_1)^{\prime}+a_0\sigma(y_1)=0
\end{eqnarray}
The last equality says that every element of $Gal(L\supset K)$ gives a
solution of (\ref{DG1}) when acting on a solution of (\ref{DG1}). Let $\{y_1,
y_2,\ldots ,y_n\}$ be the fundamental set of solutions of (\ref{DG1}). This
means that any solution of (\ref{DG1}) has the form $y=\sum_{i=1}^n \alpha_i
y_i$ where $\alpha_i$ are some constants, and
\begin{eqnarray}
\sigma(y)=\sigma\left(\sum_{i=1}^n \alpha_i y_i\right)=\sum_{i=1}^n
\alpha_i\sigma(y_i).
\end{eqnarray}
From the last equality we see that any $\sigma\in Gal(L\supset K)$ is
completely determined by its action on solutions of (\ref{DG1}), i.e.
\begin{eqnarray}
\sigma(y_i)= \sum_{i=1}^n a_{ij}y_j
\end{eqnarray}
where $a_i$ are constants. As a result we can represent the differential
Galois group as a subgroup of $GL(n,C_K)$ where $C_K$ - the field of
constants of $K$. In fact this is an algebraic subgroup of $GL(n,C_K)$
\cite{crespo07}, so in particular a Lie group. Now we can formulate the main
theorem \cite{kaplansky57}
\begin{twr}
The differential equation (\ref{DG1}) is solvable in the Liouville category
if and only if the corresponding differential Galois group is solvable.
\end{twr}
Thanks to the fact that $Gal(L\supset K)$ is a Lie group it is enough to
check whether the Lie algebra of $Gal(L\supset K)$ is solvable.

\subsection {Morales-Ramis theory - the general scheme}\label{subsec:MR}
The Morales-Ramis theory is a powerful tool for checking (non-)integrability
of Hamiltonian systems. By an integrable Hamiltonian we understand here one
which admits enough number of functionally independent, involutive with
respect to the Poisson bracket integrals of motion (this number should be
equal to the number of degrees of freedom of our system). Establishing the
nonintegrability of a Hamiltonian system is thus equivalent to proving the
nonexistence of the appropriate number of integrals of motion.

Let us shortly outline the logic of such proofs. For a general system of
nonlinear differential equations a direct count of the number of integrals of
motion is difficult - in principle there are no known methods of achieving
the goal. It is, however, clear that integrability of a nonlinear system
should be, in some way, inherited by its linearized version. Now, for linear
equations we have powerful methods based on the differential Galois theory
outlined above which can be used to relate the (non)-integrability to
properties of the differential Galois group. Reversing now the argument we
see thus that if by analyzing the differential Galois group we are able to
establish nonintegrability of the linearization, we will prove the
nonintegrability of the full, nonlinear system of equations. An additional
bonus is provided by the Hamiltonian character of the system which simplifies
the structure of the Galois group of its linearization.

To make the above outlined idea precise and workable we need some concepts
and facts. Let $(M,\omega)$ be a symplectic manifold, i.e.\ $\omega$ is a
closed ($d\omega=0$) and non-degenerate two-form on $M$. The Hamilton
equations corresponding to a Hamilton function $H$ have the form
\begin{equation}\label{DG2}
\dot{x}=X_H(x),
\end{equation}
with $X_H$ defined \textit{via} $\imath_{X_H}\omega:=\omega(X_H,\cdot)=dH$. A
linearization of (\ref{DG2}) is now achieved by defining the variational
equation (VE) along a particular non-equilibrium integral curve $\Gamma:$
$x=\phi(t)$ of (\ref{DG2}). By considering a solution of (\ref{DG2}) in the
form $x^\prime=\phi(t)+\chi(t)$ and retaining only the linear terms we obtain
the familiar result:
\begin{equation}\label{DG3}
\dot{\chi}=X_H^\prime(\phi(t))\chi.
\end{equation}
For further considerations concerning additional possible simplifications and
reductions of (\ref{DG3}) it is worth treating the derivation of it on a
slightly more formal level. We first observe that we can restrict the tangent
bundle $TM$ to $\Gamma$ obtaining the vector bundle $TM|_\Gamma$ over
$\Gamma$. We define operator $D$ on $TM|_\Gamma$ to be the Lie derivative
$\mathcal{L}_{X_H}$ restricted to $TM|_\Gamma$. More precisely to compute
$\mathcal{L}_{X_H}Y$ we extend $Y$ to $\tilde{Y}$ on a neighborhood of
$\Gamma$, compute $\mathcal{L}_{X_H}\tilde{Y}$, and restrict the result to
$\Gamma$. Operator $D$ inherits the properties of the Lie derivative, in
particular $D(fY)=f^\prime Y+fD(Y)$. The variational equation along $\Gamma$
is simply $DY=0$. It can be easily checked that choosing a suitable basis in
$TM|_\Gamma$ it can be written in the form (\ref{DG3}).

Now we want to explore the above mentioned idea that if the system
(\ref{DG2}) is integrable, then the system (\ref{DG3}) is integrable as well.
To this end we use the Ziglin lemma \cite{audin08} stating that with every
first integral $f$ of the system (\ref{DG2}) we can associate  a first
integral $f^o$ of (\ref{DG3}). Moreover if $f_1,f_2,\ldots,f_k$ are
involutive, functionally independent first integrals of (\ref{DG2}) then the
corresponding functions $f_1^o,\ldots,f_k^o$ are involutive, functionally
independent first integrals of (\ref{DG3}). The Morales-Ramis' idea was to
investigate which restrictions on the differential Galois group of the
variational equation are imposed by the complete integrability of the system
(\ref{DG2}). It turns out \cite{morales99,audin08} that:
\begin{twr}
Assume that a complex analytic Hamiltonian system is integrable in the
meromorphic function category, then the identity component of the
differential Galois group of the corresponding variational equation is
abelian.
\end{twr}
The complete integrability indeed imposes a very strong condition on the
differential Galois group of VE. To use this theorem effectively we notice
that if we know $k$ involutive integrals of motion of the system (\ref{DG2})
we can reduce the dimensionality of the system -- this is a familiar
procedure known from standard classical mechanics. More precisely the
involutive integrals of motion determine $k$ commuting Hamiltonian vector
fields which, in turn, define an isotropic\footnote{A subspace $W$ of a
symplectic space $V$ is isotropic if and only if $W\subset W^{\bot_\omega}$,
i.e. $W$ is a subspace of its orthogonal complement in the sense of the
symplectic form $\omega$.} subbundle $F$ of $TM$. We can now perform a
symplectic reduction. We have a well defined symplectic form on
$F^{\bot_\omega}/F=F^N$ and, further, we can restrict the operator $D$ to
$F^N$ because $DY\in F^{\bot_\omega}$ if $Y\in F^{\bot_\omega}$. After the
reduction we obtain from (\ref{DG3}) a $2(n-k)$ - dimensional system called
the normal variational equation (NVE) \cite{morales99,audin08}. It is proved
in \cite{morales99} that if the differential Galois group of VE is abelian
then this also true for the differential Galois group of NVE. Moreover if
there is only one missing first integral of motion then VE can be reduced to
a 2-dimensional linear matrix differential equation equivalent to some second
order linear differential one. Now the investigations of the Galois group can
be performed in an algorithmic way. From the general theory (see
Section~\ref{subsec:DG}) we know that the Galois group is a subgroup of
$GL(2,\mathbb{C})$. In fact we can restrict our search to subgroups of
$SL(2,\mathbb{C})$ \cite{kaplansky57}. Indeed, by a change of the dependent
variable $z(t)=\exp\left(\frac{1}{2}A(t)\right)y(t)$, with
$A^\prime(t)=a_1(t)$ we can eliminate the first-derivative term from the
equation $y^{\prime\prime}+a_1y^\prime+a_0y=0$ obtaining
$z^{\prime\prime}+b_0z=0$ with $b_0=a_0 - \frac{1}{4} a_0^2 - \frac{1}{2}
a_0^\prime$ without spoiling such properties of the coefficients like
meromorphicity or rationality. It is a matter of a short calculation to show
that the Wronskian, $W=z_1z_2^\prime-z_1^\prime z_2$, of two solutions $z_1$,
$z_2$ of the new equation is a constant function, a non-zero one if $z_1$ and
$z_2$ are independent. Hence, from the definition of the differential Galois
group we have $\sigma(W)=W$ for its arbitrary element $\sigma$. On the other
hand, by a straightforward calculation, $\sigma(W)=\det(\sigma)W$, hence
$\det(\sigma)=1$.

Among four possibilities allowed \cite{kaplansky57}, i.e.\ the differential
Galois group being
\begin{enumerate}
\item a finite group: the tetrahedral group, the octahedral group or the
    icosahedral group,
\item the group of matrices conjugated to the subgroup
$$\left\{ \left[
\begin{array}{cc}
 c &   0    \\
 0 & c^{-1} \\
\end{array}
\right], \left[
\begin{array}{cc}
   0    & c \\
 c^{-1} & 0 \\
\end{array}
\right],\ \ 0\ne c\in\mathbb{C} \right\}$$
i.e.\ matrices of the form
$AXA^{-1}$, where $A$ is a fixed element of $SL(2,\mathbb{C})$ and $X$
varies over the subgroup,
\item the group of triangulizable matrices,
i.e.\ matrices conjugate to the subgroup of triangular matrices, $\left\{
\left[\begin{array}{cc}
 c &   d    \\
 0 & c^{-1} \\
\end{array}
\right]\right\}$, and
\item the whole $SL(2,\mathbb{C})$,
\end{enumerate}
only in the last case it is not solvable. A practical way of
 establishing the relevant case for a particular equation is provided by
 the Kovacic algorithm \cite{kovacic86} which can be used to determine
 the differential Galois group upon analyzing poles of the coefficient
 $b_0$. Of course, if the group is not solvable then it is not abelian
 either, hence our system is not integrable.

\section{Classical and quantum chaos of $\mathfrak{su}_3$ systems}
\label{sec:qchaos}

In \cite{ghk00} a special class of systems having a compact phase space on
the classical level and, consequently, a finite-dimensional Hilbert space in
the quantum setting, was investigated. The classical limit is approached by
increasing the dimension of the Hilbert space.

As an example let us consider a collection of $N$ atoms interacting
resonantly with the electromagnetic radiation. Usually, due to imposed
resonance conditions, only a finite number $n$ of energy levels of each atom
is involved in the interaction. Transitions from the level $\ket{l}$ to the
level $\ket{k}$ of a single atom are described by the operators
$s_{kl}=\kb{l}{k}$, which span the defining representation of the Lie algebra
$\mathfrak{gl}_n(\mathbb{C})$ in the $n$-dimensional Hilbert space spanned by
the states of an atom. In a system of $N$ atoms confined to a small volume in
which they feel the same field amplitude the transitions are described by the
operators $S_{kl} =\sum_{\alpha=1}^N s_{kl}^\alpha$, where $s_{kl}^\alpha$
acts as $s_{kl}$ on the levels of the $\alpha$-th atom and as the identity on
the rest. Clearly $S_{kl}$ span the $N$-th tensor power of the defining
representation of $\mathfrak{gl}_n(\mathbb{C})$ relevant for a single atom.
The resulting representation is clearly reducible. Various preparation of the
initial state of the whole system of atoms determine its relevant irreducible
components.

 Typically the number of atoms is conserved, so $\hat N=\sum_{i=1}^NS_{ii}$ is
a constant of motion, and we can restrict the considerations to
$\mathfrak{sl}_n(\mathbb{C})$. The observables of the considered model are
constructed as polynomials in the generators. Since they have to be hermitian
we finally focus our attention on the $\mathfrak{g}=\mathfrak{su}_n$ algebra
and $G=SU_n$ Lie group.

The dynamics of the observables is governed by Heisenberg equations of motion
generated by the Hamilton operator of the considered system. The classical
limit becomes relevant when we increase the number of atoms and ask questions
about such quantities like e.g.\ energy or polarization per one atom.
Formally it consists of putting $N\rightarrow\infty$. We expect that in the
limit the generators $S_{kl}$, after appropriate scaling (e.g.\ by the number
of atoms), are mapped into classical functions on appropriate phase space in
such a way that the Heisenberg equations of motion are mapped to classical
Hamilton equations. In this way the `Dirac quantization' procedure requesting
correspondence between commutators of observables and Poisson brackets of the
corresponding phase-space functions (`classical observables') is observed.

Increasing the number of atoms $N$ results in the growing dimension of the
largest irreducible component of the constructed representation. The
construction of the classical phase space is achieved by the following
limiting procedure. For a irreducible representation of $G$ in a vector space
$V$ we consider its projective variant i.e.\ the action of the group on the
projective space $\mathbb{P}(V)$ given by $g\cdot[v]=[g\cdot v]$ for $g\in
G$, $v\in V$ and $[v]\in \mathbb{P}(V)$ -- the ray through $v$. It is known
\cite{guillemin84} that the orbit of $G$ through the point $[v]\in
\mathbb{P}(V)$ corresponding to the highest weight vector, i.e.\ the common
eigenvector of all $S_{ii}$ annihilated by all $S_{ij}$ with $i<j$, is
endowed with a natural symplectic structure. The projective orbits through
the highest weight vectors can be mapped on the orbits of the coadjoint
representation of $G$ \cite{kirillov04} i.e.\ the representation of $G$ on
the dual space $\mathfrak{g}^\ast$ of the Lie algebra $\mathfrak{g}$ with the
corresponding symplectic structure  known as the Kirillov-Kostant-Souriau
form. Coadjoint orbits are thus good candidates for classical phase-spaces.
Each irreducible representation and each above defined orbit is uniquely
determined by the highest weight vector $v$ or, equivalently, by the
corresponding eigenvalues of $S_{ii}$, the number of which equals to the rank
of the group. Alternatively and equivalently, to identify an irreducible
representation we may use independent Casimir invariants -- elements of the
enveloping algebra of $\mathfrak{g}$. From the definition they commute with
all elements of $\mathfrak{g}$, hence for each irreducible representations
they are constant multiples of the identity operators. The values of these
constants identify a representation and consequently also an orbit.

The symplectic structure on coadjoint orbits can be also obtained from a
natural Poisson structure on the linear space $\mathfrak{g}^\ast$, so called
Lie-Poisson structure (see next Section). Its symplectic leaves, i.e.\
manifolds on which the Poisson bracket determines a non-degenerate two-form,
are exactly the orbits of the coadjoint representation. Finally thus we can
identify the classical phase space with a symplectic leaf of the Lie-Poisson
structure on $\mathfrak{g}^\ast$.

For each element of the sequence of irreducible representations with growing
dimensions we obtain thus a unique well defined classical phase space. After
an appropriate scaling by the volume of the orbit \cite{schafer07} we obtain
in the limit the desired phase space of the limiting classical system,
dynamics of which is connected with the quantum one \textit{via} Dirac's
correspondence. The general idea of this construction goes back to Simon
\cite{simon80}; see \cite{sk06,schafer07} for the setting relevant for the
present considerations.

The above presented construction of the classical limit is purely
geometrical. To make it more appealing form the physical point of view let us
observe that the final result can be also obtained by treating expectation
values of the quantum observables as classical phase-space functions in the
limit of vanishing Planck constant \cite{gk98,ghk00}. The relevant
expectation values are calculated for appropriate coherent states of the
group $G$ -- these are the states `most classical' from the point of view of
uncertainty principle, hence the best approximations to the classical
description of a system \cite{perelomov86}.

The symplectic leaves of the Lie-Poisson structure may have different
dimensions. In the construction outlined above the dimensionality of the
resulting phase space may thus depend on the chosen way through the sequence
of irreducible representations. In the simplest non-trivial case of
three-level atoms the algebra of observables is spanned by eight quantities,
the generators of the Lie group $SU_3$. Irreducible representations of $SU_3$
are indexed by two independent quantum numbers -- the weight of the
highest-weight vector \cite{georgi99} -- which determine also the dimension
of the representation. In effect there exist two inequivalent ways to the
classical limit resulting in a six- (in a generic case) or four- (in a
degenerate case) dimensional classical phase space. This observation was a
basic point of the paper \cite{ghk00}, where it was shown that the
dimensionality of the classical space determines not only integrability
properties of a specific class of classical Hamilton functions, but also some
statistical properties of spectra of the quantum Hamiltonians for which the
classical system in question is the classical limit outlined above. Both
investigations of spectra on the quantum level and integrability properties
on a classical one were performed numerically. We are now in position to
prove analytically the non-integrability for a concrete member of the
considered class.

The Hamiltonian we consider is quadratic in the generators $S_{ij}$,
\begin{equation}\label{qH}
\widehat{H}=3(S_{12}^2+S_{21}^2)+15(S_{13}S_{32}+S_{23}S_{31})
\end{equation}
(see remarks in \cite{gk98,ghk00} for the possibilities of experimental
realizations). It is easy to show that $[H,Y]=0$, where
$\widehat{Y}=S_{11}+S_{22}-2S_{33}$. The commutation relation survives the
classical limit providing thus an integral of motion. As a result we obtain a
classical Hamiltonian system with the Hamilton function
\begin{equation}\label{cH0}
H=3(s_{12}^2+s_{21}^2)+15(s_{13}s_{32}+s_{23}s_{31}),
\end{equation}
admitting an integral of motion
\begin{equation}\label{cY0}
Y=s_{11}+s_{22}-2s_{33},
\end{equation}
where $s_{ij}$ are coordinates on $\mathfrak{su}_3^\ast$ dual to $S_{ij}$.

In the next section we explain how to obtain from (\ref{cH0}) the
corresponding Hamilton equations of motion on $\mathfrak{su}_3^\ast$, in
particular we describe the above mentioned Lie-Poisson structure. Since the
classical phase space is either four or six dimensional the classical system
(\ref{cH0}) is integrable in the former and possibly non-integrable (if there
are no other unknown integrals of motion) in the latter case.

\section {Lie-Poisson structure on $\mathfrak{su}_3^\ast$}\label{sec:LP}
As stated in the Introduction and explained in the preceding section, we are
interested in a class of systems which are obtained as the classical limits
of some quantum systems with $SU(3)$ symmetry \cite{gk98,ghk00,sk06}, and the
classical limit of such a system can be considered as a Hamiltonian system on
$\mathfrak{su}_3^\ast$. It is instructive to consider the problem in a
slightly more general setting where $\mathfrak{su}_3$ is substituted by an
arbitrary Lie algebra.

Let $(\mathfrak{g},[\cdot,\cdot])$ be an arbitrary Lie algebra. We can equip
its dual space $\mathfrak{g}^\ast$ with the canonical Lie-Poisson structure
given by the Poisson bracket,
\begin{equation}\label{LP}
\{f,g\}(x)=\langle x,[(df)_x,(dg)_x]\rangle,
\end{equation}
where $f,g \in C^\infty(\mathfrak{g}^\ast)$ and $\langle,\rangle$ denotes the
pairing between $\mathfrak{g}^\ast$ and $\mathfrak{g}$.  The bracket
(\ref{LP}) is well defined because $(df)_x:\mathfrak{g}^\ast\rightarrow R$ is
an element of $\mathfrak{g}^{\ast\ast}$ and in the case of finite dimensional
vector spaces we have $\mathfrak{g}^{\ast\ast}=\mathfrak{g}$. The bracket
defined in this way is of course bilinear and antisymmetric. It is also a
differentiation i.e.\ it satisfies the Leibniz rule. It is convenient to
describe the Lie-Poisson structure on $\mathfrak{g}^\ast$ in terms of a
Poisson bivector,
\begin{equation}\label{biv}
\eta=c_{ij}^k x_k\frac{\partial}{\partial x_i}\wedge\frac{\partial}
{\partial x_j},
\end{equation}
where $c_{ij}^k$ are the structure constants of $\mathfrak{g}$ corresponding
to some basis $e_1,...,e_n$, i.e.\ $[e_i,e_j]=c_{ij}^ke_k$, and coordinates
$x_1,...,x_n$ are the vectors $e_1,...,e_n$ considered as linear functions on
$\mathfrak{g}^\ast$. To show that the bivector induced by Poisson bracket
(\ref{LP}) is the same as the one given by Eq.~(\ref{biv}), it is enough to
check it on the linear functions. After completing this easy taks we may thus
write the Hamilton equations of motion for na arbitrary function $f$ on
$\mathfrak{g}^\ast$,
\begin{equation}\label{chem}
\frac{df}{dt}=\{H,f\}=\eta(H,f)
\end{equation}

The Poisson bivector $\eta$ is degenerate on $\mathfrak{g}^\ast$ due to the
existence of Casimir functions which have a vanishing Poisson bracket with
any function. Thus we do not obtain directly any symplectic structure on
$\mathfrak{g}^\ast$. On the other hand when we restrict $\eta$ to its
symplectic leaves determined by constant values of independent Casimir
functions we end up with well defined symplectic manifolds. Indeed, $\eta$
defines a morphism
\begin{equation}\label{etasharp}
\eta^\sharp:T\mathfrak{g}^{\ast\ast}\rightarrow T\mathfrak{g}^\ast,
\quad df|_x\mapsto X_f|_x,
\end{equation}
where $X_f=\{f,\cdot\}$. It generates a distribution $\mathcal{D}=\bigcup _x
D_x$, $x\in M$, and $D_x=image(\eta^\sharp_x)$. This distribution is
involutive, (i.e.\ $[X,Y]\in\mathcal{D}$, for $X,Y\in\mathcal{D}$), hence
from the Frobenius theorem $\mathcal{D}$ is tangent to some generalized
foliation $\mathcal{F}$. The restriction of $\eta$ to leaves,
$\mathcal{F}_x$, $x\in M$, of the foliation $\mathcal{F}$ is a well defined,
non-degenerate Poisson bivector, so it defines a symplectic structure on
$F_x$. The symplectic leaves $\mathcal{F}_x$ are exactly the coadjoint orbits
of $G$, and the corresponding symplectic form providing a symplectic
structure is the announced Kirillov-Kostant-Souriau one \cite{kirillov04}.
Its explicit form can be easily deduced from the definition of $\eta$,  It
is, however, often easier to work with the Poisson structure (\ref{LP}) on
the whole $\mathfrak{g}$ than with its restriction to leaves and treat the
Casimir functions as constants motion determining by their initial values a
manifold to which the motion is restricted. This is the way we will follow in
our case.

We may now specify the above general considerations to the $\mathfrak{su}_3$
case of the present interest. To this end we have to chose some basis
$\mathfrak{su}_3$ and find explicitly the bivector $\eta$. The basis of our
choice consists of the standard Gell-Mann matrices (see e.g.\
\cite{georgi99}; we use them in slightly different order) multiplied by the
imaginary unit $i$,
\begin{eqnarray}\label{gellmann}
  e_1&=& \left[ \begin {array}{ccc} 0&1&0\\\noalign{\medskip}-1&0&0
\\\noalign{\medskip}0&0&0\end {array} \right], \quad
  e_2=\left[ \begin {array}{ccc} 0&i&0\\\noalign{\medskip}i&0&0
\\\noalign{\medskip}0&0&0\end {array} \right], \quad
  e_3=\left[ \begin {array}{ccc} i&0&0\\\noalign{\medskip}0&-i&0
\\\noalign{\medskip}0&0&0\end {array} \right], \nonumber
 \\
  e_4&=& \left[ \begin {array}{ccc} 0&0&1\\\noalign{\medskip}0&0&0
\\\noalign{\medskip}-1&0&0\end {array} \right], \quad
 e_5=\left[ \begin {array}{ccc} 0&0&i\\\noalign{\medskip}0&0&0
\\\noalign{\medskip}i&0&0\end {array} \right], \quad
 e_6=\left[ \begin {array}{ccc} 0&0&0\\\noalign{\medskip}0&0&1
\\\noalign{\medskip}0&-1&0\end {array} \right],
\\
 e_7 &=&\left[ \begin {array}{ccc} 0&0&0\\\noalign{\medskip}0&0&i
\\\noalign{\medskip}0&i&0\end {array} \right], \quad
 e_8=\frac{1}{\sqrt{3}}\left[ \begin {array}{ccc} i&0&0\\\noalign{\medskip}0&i&0
\\\noalign{\medskip}0&0&-2i\end {array} \right]. \nonumber
\end{eqnarray}
In this basis we have $\eta=c_{ij}^k x_k\frac{\partial}{\partial
x_i}\wedge\frac{\partial}{\partial x_j}=\eta_{ij}\frac{\partial}{\partial
x_i}\wedge\frac{\partial}{\partial x_j}$ with the coefficients $\eta_{ij}$
given as
\begin{displaymath}
\begin{array}{llll}
\eta_{12}=2x_3,  & \eta_{25}=-x_6,  & \eta_{46}=-x_1,
& \eta_{78}=\sqrt{3}x_6.\\
\eta_{13}=-2x_2,  & \eta_{26}=x_5,  & \eta_{47}=x_2,  &\\
\eta_{14}=-x_6,  & \eta_{27}=-x_4,  & \eta_{48}=-\sqrt{3}x_5,&  \\
\eta_{15}=-x_7,  & \eta_{34}=x_5,  & \eta_{56}= -x_2,  &\\
\eta_{16}=x_4,  & \eta_{35}=-x_4,  & \eta_{57}=-x_1,  &\\
\eta_{17}=x_5, & \eta_{36}=-x_7, & \eta_{58}= \sqrt{3} x_4, &\\
\eta_{23}=2x_1,  & \eta_{37}=x_6,  & \eta_{67}= (-x_3+\sqrt{3} x_8),& \\
\eta_{24}=x_7,  & \eta_{45}=(x_3+\sqrt{3}x_8),  & \eta_{68}= -\sqrt{3} x_7,
& \\
\end{array}
\end{displaymath}

The bivector $\eta$ is degenerate and there are two functionally independent
Casimir functions $c_1$, $c_2$, ($\eta^\sharp(dc_1)=0=\eta^\sharp(dc_2)$),
given by $c_1=\alpha \tr(X^2)$ and $c_2=\beta\tr(X^3)$, where $X$ is a
generic matrix belonging to $\mathfrak{su}_3$ algebra,
\begin{displaymath}\label{gensl2c}
X=\left(
\begin{array}{ccc}
ix_3+i\frac{x_8}{\sqrt{3}}&x_1+ix_2&x_4+ix_5\\
-x_1+ix_2&-ix_3+i\frac{x_8}{\sqrt{3}}&x_6+ix_7\\
-x_4+ix_5&-x_6+x_7&-2i\frac{x_8}{\sqrt{3}}\\
\end{array}\right),
\end{displaymath}
and $\alpha$, $\beta$ are arbitrary constants. To keep the consistency with
\cite{ghk00} where hermitian rather than antihermitian matrices were used to
represent $\mathfrak{su}_3$ algebra, we choose $\alpha=-1$, $\beta=i$.

\section{The non-integrability proof}

As we noticed in Section~\ref{sec:LP}, the $\mathfrak{su}_3^\ast$ Lie-Poisson
structure has two Casimir functions $c_1$ and $c_2$, hence the dimension of a
generic leaf is six, but there are cases for which it reduces to four
\cite{gk98}. It is interesting to check whether a Hamiltonian system defined
on the whole $\mathfrak{su}_3^\ast$ which possess an additional first
integral besides the Hamilton function itself, `placed' on leaves of
different dimension is integrable or not. By `placed' we mean that initial
conditions determine the leaf on which the time evolution takes place. Of
course such a Hamiltonian system is integrable on four dimensional leaves
(there are two first integra    ls in involution), but in the case of six
dimensional leaves an additional first integral needed for integrability may
lack. Our aim is to prove that for some particular polynomial Hamiltonian
systems an additional first integral is indeed missing, with (\ref{cH0})
treated as a concrete example.

Hamilton functions $H$ on $\mathfrak{su}_3^\ast $ we are interested in are
given as second order polynomials in the $x_i$ coordinates given in
Section~\ref{sec:LP}. The particular example given by (\ref{cH0}) takes in
the new variables the form\footnote{The change of variables from $s_{ij}$ to
$x_i$ can be easily red off from (\ref{gensl2c}) \textit{via} $X_{ij}=s_{ij}$
},
\begin{equation}\label{cH1}
H=6(x_1^2-x_2^2-5 x_4 x_6-5 x_5 x_7).
\end{equation}

The resulting Hamilton equations are given by:
\begin{equation}\label{system}
\frac{dx}{dt}=\eta^\sharp(dH).
\end{equation}

In the coordinates $x_i$ they form a set of eight differential equations:
\begin{eqnarray}\label{eqs-full}
\frac{dx_1}{dt}&=& 24 x_2 x_3+30 x_4^2+30 x_5^2-30 x_6^2-30 x_7^2,\nonumber\\
\frac{dx_1}{dt}&=&24 x_1 x_3,\nonumber\\
\frac{dx_3}{dt}&=&-48 x_1 x_2+60 x_5 x_6-60 x_4 x_7, \nonumber\\
\frac{dx_4}{dt}&=& -30 x_1 x_4+30 x_2 x_5-12 x_1 x_6-12 x_2 x_7+30 x_7 (x_3+\sqrt{3 x_8}),\nonumber\\
\frac{dx_5}{dt}&=& -30 x_2 x_4-30 x_1 x_5+12 x_2 x_6-12 x_1 x_7+30 x_6 (-x_3-\sqrt{3}x_8),\nonumber\\
\frac{dx_6}{dt}&=&12 x_1 x_4-12 x_2 x_5+30 x_1 x_6+30 x_2 x_7+30 x_5 (-x_3+\sqrt{3}x_8),\nonumber\\
\frac{dx_7}{dt}&=& 12 x_2 x_4+12 x_1 x_5-30 x_2 x_6+30 x_1 x_7+30 x_4 (x_3-\sqrt{3}x_8),\nonumber\\
\frac{dx_8}{dt}&=& 0.
\end{eqnarray}
The last equations reflects the fact that (\ref{cY0}) is a constant of
motion, since in the new coordinates we have simply $Y=x_8$. As we announced
we are working in the full $\mathfrak{su}^\ast$ space, we know thus two
additional integrals of motion given by the Casimir functions $c_1$ and
$c_2$. In \cite{ghk00} it was shown that the for the classical limit of
$SU(3)$ systems obtained \textit{via} the procedure outlined in
Section~\ref{sec:qchaos}, the values of the Casimir constants of motion can
be parameterized by a single number $q\in[0,1]$,
\begin{eqnarray}\label{cas-q}
c_1&=&\frac{2}{3}(q^2-q+1), \nonumber \\
c_2&=&\frac{1}{9}(-2q^3+3q^2+3q-2).
\end{eqnarray}
It was also shown in \cite{ghk00} that for $q=0$ and $q=1$ the leaf on which
the system evolves is four dimensional whereas for $q\in]0,1[$ the leaves are
six dimensional.

Fixing the values of $c_1$ and $c_2$ by choosing a particular $q$ restricts
also possible values of $x_8$. A possible compatible choice is
\begin{equation}\label{x8-q}
x_8=\frac{\sqrt{3}}{2}(1-2q).
\end{equation}

To use Morales-Ramis theory we have to find some particular, as simple as
possible, non-equilibrium solution. If we put
\begin{equation}\label{part}
x_1=x_2=x_5=x_6=x_8=0, \quad x_4=x_7,
\end{equation}
the system (\ref{eqs-full}) reduces to two differential equations:
\begin{eqnarray}\label{eqs-2d}
\frac{dx_3}{dt}&=&-60 x_4^2, \\ \nonumber
\frac{dx_4}{dt}&=&30 x_3x_4.\\\nonumber
\end{eqnarray}
The system can be easily solved since we know two constant Casimir functions.
We want to choose such values of them that the corresponding symplectic leaf
is six-dimensional. For our choice (\ref{part}) the Casimir functions
simplify to $c_1=2(x_3^2+2x_4^2)$ and $c_2=0$. Observe that a choice
$q=\frac{1}{2}$ which, as stated above corresponds to a six-dimensional phase
space, indeed gives $c_2=0$ [see (\ref{cas-q})], forcing in addition $x_8=0$
[see (\ref{x8-q})]. We will prove that in the case $q=\frac{1}{2}$ the whole
system (\ref{eqs-full}) is not integrable in the Liouville sense.

The solution of (\ref{eqs-2d}) for $q=\frac{1}{2}$ is found to be
\begin{eqnarray}\label{sol-2d}
x_3=\frac{1}{2}\tanh(-15t)\qquad
x_4=\frac{\sqrt{2}}{4}\sqrt{1-\tanh(-15t)},
\end{eqnarray}
It defines a particular integral curve $\Gamma$ of the full system. We can
now find variational equation along the obtained solution,
\begin{equation}\label{vareqs}
\frac{dy}{dt}=A(t)y,
\end{equation}
where the matrix $A(t)$ is given by \small
\begin{displaymath}
A(t)=\left(
\begin{array}{cccccccc}
 0 & 24 x_3& 0 & 60 x_4& 0 & 0 & -60 x_4& 0 \\
 24 x_3& 0 & 0 & 0 & 0 & 0 & 0 & 0 \\
 0 & 0 & 0 & -60 x_4& 0 & 0 & -60 x_4& 0 \\
 -30 x_4& -12 x_4& 30 x_4& 0 & 0 & 0 & 30 x_3& 30 \sqrt{3} x_4\\
 -12 x_4& -30 x_4& 0 & 0 & 0 & -30 x_3& 0 & 0 \\
 12 x_4& 30 x_4& 0 & 0 & -30 x_3& 0 & 0 & 0 \\
 30 x_4& 12 x_4& 30 x_4& 30 x_3& 0 & 0 & 0 & -30 \sqrt{3} x_4\\
 0 & 0 & 0 & 0 & 0 & 0 & 0 & 0
\end{array}
\right).
\end{displaymath}
\normalsize The tangent space at each point $x$ of curve $\Gamma $ is $D_x$
where $D_x=image(\eta^\sharp_x)$. It can be described effectively using
Casimir functions, namely $V_x\in T_xF_x$ if and only if $dc_1(V_x)=0$ and
$dc_2(V_x)=0$. If we denote by $V_x=\{V_{x}^1,\ldots,V_{x}^8\}$ the
coefficients of $V_x$ with respect to the basis corresponding to $x_1,\ldots,
x_8$, the conditions $dc_1(V_x)=0$ and $dc_2(V_x)=0$ take the form:
\begin{eqnarray}
x_3V_x^3+x_4(V_x^4+V_x^7)&=&0, \nonumber\\
3x_4^2V_x^1+3 x3 +x_4(V_x^4-V_x^7)+\sqrt{3}(x_3^2 -  x_4^2)V_x^8&=&0
\end{eqnarray}
Using these relations we notice that $V_x^3$ and $V_x^8$ are completely
determined if we know $V_{x}^1,V_x^4,V_x^7$, so we can reduce (\ref{vareqs})
to a set of six differential equations
\begin{equation}\label{varex1}
\frac{d\xi}{dt}=B(t)\xi,
\end{equation}
where the matrix $B(t)$ is given by: \small
\begin{displaymath}
B(t)=\left(\begin{array}{cccccc}
 0 & 24 x_3& 60 x_4& 0 & 0 & -60 x_4\\
 24 x_3& 0 & 0 & 0 & 0 & 0 \\
 \frac{30x_4 (x_3^2 +2x_4^2)}{-x_3^2+x_4^2}& -12 x_4& \frac{30 x_4^2 (-4  x_3^2+x_4^2)}{x_3^3-x_3
x_4^2}& 0 & 0 & \frac{30(x_3^4+x_3^2 x_4^2+x_4^4)}{x_3^3-x_3 x_4^2}\\
 -12 x_4& -30 x_4& 0 & 0 & -30 x_3& 0 \\
 12 x_4& 30 x_4& 0 & -30 x_3& 0 & 0 \\
 30x_4+\frac{90x_4^3}{x_3^2-x_4^2}& 12 x_4& \frac{30 (x_3^4+x_3^2 x_4^2+x_4^4)}{x_3^3-x_3
x_4^2}& 0 & 0 & \frac{30 x_4^2 (-4 x_3^2+x_4^2)}{x_3^3-x_3 x_4^2}
\end{array}
\right).
\end{displaymath}
\normalsize The set of differential equations (\ref{varex1}) can be reduced
to a normal variational equation ($NVE$) using the fact that
$X_H=\eta^\sharp(dH)$ and $X_8=\eta^\sharp(dx_8)$ are solutions of
(\ref{varex1}). To perform the reduction we need the symplectic form at each
point of the curve $\Gamma$. It can be obtained by inversion of $\eta_\Gamma$
which is the restriction of $\eta$ to the curve $\Gamma$. It is possible
since $\eta$ is non-degenerate on $\Gamma$ ($\Gamma\subset F_x$ for some
$x$). Explicit calculations give the symplectic form $\omega_\Gamma$ along
$\Gamma$ as $\omega_\Gamma=\omega_{ij}dx_i\wedge dx_j$, where $i$ and $j$
belong to $\{1,2,4,5,6,7\}$ and the coefficients $\omega _{ij}$ read:
\begin{displaymath}
\begin{array}{llll}
\omega_{12}=-\frac{x_3}{2(x_3^2-x_4^2)}  & \omega_{27}=\frac{x_4}{2(x_4^2-x_3^2)}  \\
\omega_{15}=\frac{x_4}{2(x_4^2-x_3^2)}  & \omega_{45}=\frac{x_4^2-2x_3^2}{2x_3(x_3^2-x_4^2)} \\
\omega_{16}=\frac{x_4}{2(x_3^2-x_4^2)}  & \omega_{46}=\frac{x_4^2}{2x_3(x_3^2-x_4^2)} \\
\omega_{24}=\frac{x_4}{2(x_3^2-x_4^2)}  & \omega_{57}=\frac{-x_4^2}{2x_3(x_3^2-x_4^2) } \\
\omega_{45}=\frac{2x_3^2-x_4^2}{2x_3(x_3^2-x_4^2)}& \\
\end{array}
\end{displaymath}
The key point now is to find a symplectic basis including $X_H$ and $X_8$,
i.e.\ a set of six vector fields such that $\eta_\Gamma\sim X_H\wedge
\tilde{X}_H+X_8\wedge\tilde{X}_8+X\wedge\tilde{X}$. Such a basis always exist
\cite{morales99}, and in our case it is formed by
\begin{alignat}{2}\label{symp-basis}
X_H&=30\,x_3 x_4\left(\frac{\partial}{\partial x_4}+\frac{\partial}{\partial
x_7}\right), &\qquad
\tilde{X}_H&=\frac{-1}{60\,x_4}\left(\frac{\partial}{\partial
x_5}+\frac{\partial}{\partial x_6}\right), \nonumber\\
X_8&=x_4\left(\frac{\partial}{\partial x_5}-\frac{\partial}{\partial
x_6}\right),
&\qquad\tilde{X}_8&=\frac{x_3^2-x_4^2}{2\,x_3x_4}\left(\frac{\partial}
{\partial x_4}+\frac{\partial}{\partial x_7}\right), \nonumber\\
X&=-2x_3\frac{\partial}{\partial x_1}+x_4\left(\frac{\partial}{\partial
x_4}-\frac{\partial}{\partial x_7}\right),
&\qquad\tilde{X}&=-\frac{\partial}{\partial
x_2}+\frac{x_4}{2x_3}\left(\frac{\partial}{\partial
x_5}-\frac{\partial}{\partial x_6}\right).
\end{alignat}
We can express the equation (\ref{varex1}) in the basis (\ref{symp-basis})
as:
\begin{equation}
\frac{d\chi}{dt}=P^{-1}(B(t)P-\dot{P})=C(t)\chi
\end{equation}
where $\chi=P\xi$ and $P$ is the change of basis matrix:
\begin{displaymath}
P(t)=\left(\begin{array}{cccccc}
 0 & 0 & 0& 0  & -2x_3&0\\
 0& 0 & 0 & 0 & 0 & -1 \\
 30x_3x_4&0&0& \frac{x_3^2-x_4^2}{2x_3x_4}& x_4 & 0  \\
 0& x_4 & \frac{-1}{60x_4} & 0 & 0 & \frac{x_4}{2x_3} \\
  0& -x_4 & \frac{-1}{60x_4} & 0 & 0 & -\frac{x_4}{2x_3} \\
 30x_3x_4&0&0& \frac{x_4^2-x_3^2}{2x_3x_4}& -x_4 & 0  \\
 \end{array}
 \right)
\end{displaymath}
The derivative matrix $\dot{P}$ can be easily computed using (\ref{eqs-2d}).
The final result reads as
\begin{displaymath}
C(t)=\left(\begin{array}{cccccc}
 0 & 0 & 0& 0  & 0&0\\
 0& 0 & 0 & 0 & 0 & 30-\frac{30x_4^2}{x_3^2} \\
0 & 0 & 0& 0  & 0&0\\
0 & 0 & 0& 0  & 0&0\\
0 & 0 & 0& -30+\frac{30x_4^2}{x_3^2}  & 0&12\\
0 & 0 & 0& 0  & 48x_3^2&0\\
 \end{array}
 \right)
\end{displaymath}
Thus the NVE is a simple $2\times2$ matrix differential equation:
\begin{displaymath}
\frac{d\tilde{\chi}}{dt}=\left(\begin{array}{cc}
0&12\\
48x_3^2&0\\
\end{array}
\right)\tilde{\chi}.
\end{displaymath}
Writing
\begin{equation}\label{chicoord}
\tilde{\chi}=\left[\begin{array}{c}
 \tilde{\chi}_1 \\
 \tilde{\chi}_2 \\
\end{array}\right]
\end{equation}

we find the corresponding second order differential equation,
\begin{equation}
\frac{d^2\tilde{\chi}_1}{dt^2}-576x_3^2\tilde{\chi}_1=0.
\end{equation}
Making use of (\ref{sol-2d}) we get:
\begin{equation}
\frac{d^2\tilde{\chi}_1}{dt^2}-144\tanh^2(-15t)\tilde{\chi}_1=0
\end{equation}
The substitution $y=\tanh(-15t)$ transforms the equation to:
\begin{equation}\label{final}
255\left(1-y^2\right)^2\frac{d^2 f}{dy^2}-450\left(1-y^2\right)y\frac{df}{dy}-144y^2f=0
\end{equation}
The equation is fully prepared to be treated by the Kovacic algorithm. We are
not going to describe it here (see the references \cite{kovacic86} and
\cite{morales99} for details). The algorithm produces a solution if an
equation is integrable in the Liouville sense, and, what is more important
for us, it determines as a byproduct the differential Galois group
identifying it as one among those listed at the end of
Section~\ref{subsec:MR}. For our equation (\ref{final}) the result is that
the Galois is not solvable. It is thus not abelian and the Hamiltonian
(\ref{cH1}) is not integrable in the meromorphic function category on the
chosen, six-dimensional phase space. To check the Liouville integrability in
a concrete case like (\ref{final}) one can also use an implementation of the
Kovacic algorithm in symbolic manipulation programs, e.g.\
\textit{kovacicsols} from \textit{Maple 12} which returns a list of
Liouvillian solutions if they exist and the empty set in the opposite case.
The occurrence of the latter case is thus a proof of the nonintegrability of
the full system.

\section{Conclusions and outlook}
We proved in an analytical way the non-integrability of a specific quadratic
Hamilton function defined on the Lie algebra $\mathfrak{su}_3$ obtained as a
classical limit of a quantum Hamiltonian. The motivation was thoroughly
presented in the Introduction and Section~\ref{sec:qchaos}, here we want to
conclude that the achieved result fills, at least partially, a gap in the
reasoning of \cite{ghk00}.

We would like also to highlight some novelties of our investigation. Since
the birth of the Morales-Ramis theory there has been many successful attempts
to apply it to concrete physical situations
\cite{saenz00,almeida04,maciejewski04a,maciejewski05a}. The investigated
system were, usually, of the standard type with Hamilton function of the form
of a sum of the kinetic and potential energies, the kinetic energy being a
quadratic form in the canonical\footnote{The canonical variables, say $p$ and
$q$ are those in which the symplectic form reads (at least locally)
$\omega=\sum_i dp_i\wedge dq^i$} momentum variables, defined in a
topologically simple phase space. This is not the case for the Hamilton
function treated in our paper. It is a quadratic polynomial in non-canonical
variables on a compact symplectic manifold.

The presented reasoning can be applied to other symplectic leaves, other
Hamilton functions on $\mathfrak{su}_3$ (or other Lie algebras). The desired
result would be a classification of such Hamilton functions with respect to
their integrability.

\section{Acknowledgments}
We gratefully acknowledge the inspiration which came from numerous
discussions with Ingolf Sch\"affer. The work was supported by SFB/TR12
'Symmetries and Universality in Mesoscopic Systems' program of the Deutsche
Forschungsgemeischaft and Polish MNiSW grant no.\ DFG-SFB/38/2007.


\end{document}